# Non-extensive entropy from incomplete knowledge of Shannon entropy?

F. Sattin
Consorzio RFX, Associazione EURATOM-Enea per la fusione
Corso Stati Uniti 4, 35127 Padova - ITALY
e-mail: fabio.sattin@igi.cnr.it

**Abstract**
*In this paper we give an interpretation of Tsallis' nonextensive statistical mechanics based upon the information-theoretic point of view of Luzzi et al. [cond-mat/0306217; cond-mat/0306247; cond-mat/0307325], suggesting Tsallis entropy to be not a fundamental concept but a derived one, stemming from an incomplete knowledge of the system, not taking properly into account its interaction with the environment. This interpretation seems to avoid some problems occuring with the original interpretation of Tsallis statistics.*

Nonextensive statistical mechanics (NESM) introduced by Tsallis [1] has been raising considerable interest for its ability in interpreting several experimental results (for a review of the applications of Tsallis formalism, see the updated website: http://tsallis.cat.cbpf.br/biblio.htm).

The standard way of introducing NESM is through the definition of a generalized entropy functional:

$$S_q = \frac{1 - \sum_i p_i^q}{q - 1} \quad , \qquad (1)$$

from which a whole generalized thermodynamics can be recovered (see, e.g. [2]).

With the usual procedure of constrained entropy maximimization, one can deduce from (1) the equilibrium Probability Distribution Function (PDF) for any given system. Indeed, it is possible to show that the set of probabilities $p_i$ maximizing $S_q$ within the canonical ensemble are

$$p_i = \frac{1}{Z_q}\left[1 - \beta(1-q)\varepsilon_i\right]^{1/(1-q)} \qquad (2)$$

with $\varepsilon_i$ energy of the *i-th* microstate and the Lagrange multiplier β that plays the role of generalized inverse temperature. This expression enjoys so widespread consideration that Tsallis' NESM has become often sinonimous for power-law PDFs. Actually, there is a huge and ever-increasing number of systems where PDFs of some quantity are measured and found to be consistent with power laws (2).

It is important to point out that Eq. (1) is a sufficient condition for (2) to hold, but by no means a necessary one. There are other different definitions for the entropy which lead to the same extremizing $p_i$-distribution (e.g., the Rényi entropy). Even more important, PDFs of the type (2) can be recovered without any explicit recourse to a definition of entropy whatsoever: they can be seen as the equilibrium or quasi-equilibrium distributions for systems interacting with thermal baths subject to fluctuations. This fact was first empirically found for a particular system by Wilk and Wlodarczyk [3], then Beck [4] provided a theoretical interpretation (see also Sattin and Salasnich [5]), pointing it out that, in principle, fluctuations of the thermal bath could lead to several classes of PDFs, of which Eq. (2) represents only a particular case. Finally, Beck and Cohen [6] enframed all these findings within the unifying concept of superstatistics.

A matter of debate is, therefore, the role that must be assigned to the generalized entropy $S_q$ (Eq. 1): has it a physical meaning just like standard Shannon entropy has in thermostatistics, or is just a useful compact expression to derive non-exponential PDFs? Indeed, some authors [7] have pointed out that $S_q$ should have a meaning just within an information-thoretic approach, as an informational entropy, in situations where the observer cannot have access to a complete characterization of the system (i.e., the states of the system cannot be properly characterized, for example systems with fractal structure, or with long range interactions).

In this work we will be providing some considerations about this question, following essentially the path of Luzzi *et al* [7]. We do not pretend to give a definite answer here; in particular we must admit the severe constraint that our calculations hold only for the nonextensive exponent close to unity. This covers many of the cases tracted in literature, but there are also several exceptions. However, we will provide the reader with some comments and analytical calculations supporting Luzzi *et al*'s viewpoint..

We will try, hence, to give Eq. (1) a different, information-theoretic interpretation: Tsallis entropy is looked like a partial information entropy, arising from incomplete information over the whole system under consideration. The starting point is the empirical observation of the existence of PDFs of the kind (2) (or, more generally, non-Gaussian), together with the explanation of their existence as due the interaction of the system $\Sigma$ with a fluctuating thermal bath *B*. From this, expressions of the form (1) can be recovered, but they do not express any fundamental property of the system. Instead, they are simply approximate expressions arising when we have an incomplete knowledge of the whole system $\Sigma+B$. We remind that this fact is perhaps implicit in Beck's superstatistics formalism, and was recently verified through numerical experiments by Potiguar and Costa [8]. These authors showed that PDFs (2) necessarily arise when our system (endowed with ordinary Maxwell-Boltzmann statistics) interacts with a thermal bath made of a finite number of degrees of freedom, hence fluctuacting in its macroscopic quantities (e.g., temperature).

In the following we will use the *unnormalized* form of Tsallis statistics. Nowadays, there is evidence in support of a *normalized* form, where mean quantities (observables) must be divided by $\sum p^q$. However, we do not expect the use of one form rather than another may affect the present results.

Let us consider the (Shannon) entropy of the total system $\Sigma + B$:

$$S = -\sum_{ij} b_i s(j|i) \ln[b_i s(j|i)] \tag{3}$$

where $b_i$ is the probability for the bath *B* to be in the *i-th* state, and $s(j|i)$ is the conditional probability for the system $\Sigma$ to be in the state *j* if *B* is in the state *i*. If $\Sigma+B$ is an isolated system, its equilibrium distribution can be found by extremizing (3) with the normalization condition $\sum_{ij} b_i s(j|i) = 1$, that is

$$\delta F \equiv \delta\left(S - \lambda \sum_{ij} b_i s(j|i)\right) = 0 \rightarrow b_i s(j|i) = \text{const} \tag{4}$$

Note that this does not imply $s$ = const, that is, is $\Sigma+B$ to follow microcanonical statistics, not $\Sigma$.

Let us now suppose that there is a large (almost but not exactly unitary) probability for the bath to be just in one single state, say $I$; thus write

$$b_i \sim \delta_{iI} + (1-\delta_{iI})(1-\alpha)d_i \quad . \tag{5}$$

In the expression above $(1-\alpha)$ is just a small parameter, useful for doing perturbative expansions. Its meaning is that of an effective width of the fluctuations around the most probable state $I$. If we insert Eq. (5) into (3) we can write

$$S \sim -\sum_j s(j|I)\ln[s(j|I)] - (1-\alpha)\sum_{i\neq I,j} d_i s(j|i)\ln[d_i s(j|i)] \quad . \tag{6}$$

and the functional $F$ defined in (4) is rewritten

$$F \sim -\sum_j s(j|I)\ln[s(j|I)] - (1-\alpha)\sum_{i\neq I,j} d_i s(j|i)\ln[d_i s(j|i)] - \lambda\sum_j s(j|I) - \lambda(1-\alpha)\sum_{i\neq I,j} d_i s(j|i) \tag{7}$$

The first term in the r.h.s of (6) would be the ordinary entropy for the system $\Sigma$, if it were isolated.

Let us now imagine a researcher wishing to determine the equilibrium PDF for $\Sigma$, ignoring that it is interacting with $B$; hence, he would assume $s(j|i) \equiv s(j)\,\forall i$, and try to extremize the microcanonical-ensemble functional

$$F_\mu = -\sum_j s(j)\ln[s(j)] - \lambda\sum_j s(j) \quad . \tag{8}$$

or, allowing for a weak energy exchange, the canonical-ensemble functional

$$F_C = -\sum_j s(j)\ln[s(j)] - \lambda\sum_j s(j) - \beta\sum_j \varepsilon_j s(j) \tag{9}$$

Of course, in both cases, he would not find agreement with experiment, since distributions extremizing (8) or (9) generally do not extremize (7). At this point, one could be tempted to modify the definition of the entropy. Actually, if we write the equivalent of (9) using Tsallis rules, we get the functional

$$F_q = S_q - \lambda\sum_i s_s - \beta\sum_i \varepsilon_i s_i^q \tag{10}$$

(Note the use of $s^q$ instead of $s$ in the Lagrange multiplier relative to energy).

Supposing $q \sim 1$ and expanding (10) to first order in $(1 - q)$, we find

$$F_q \sim -\sum_j s_j \ln(s_j) - \lambda \sum_j s_j - \beta \sum_j \varepsilon_j s_j - (1-q)\left[\frac{1}{2}\sum_j s_j \ln^2(s_j) + \beta \sum_j \varepsilon_j s_j \ln(s_j)\right] + O(1-q)^2$$

(11)

Eqns. (7) and (11) can be identified provided the following relations hold:

$$\sum_j s(j|I)\ln[s(j|I)] \leftrightarrow \sum_j s_j \ln(s_j)$$

$$\lambda \sum_j s(j|I) \leftrightarrow \lambda \sum_j s_j$$

$$\lambda(1-\alpha)\sum_{i \neq I, j} d_i s(j|i) \leftrightarrow \beta \sum_j \varepsilon_j s_j \qquad (12)$$

$$(1-\alpha)\sum_{i \neq I, j} d_i s(j|i) \ln[d_i s(j|i)] \leftrightarrow (1-q)\sum_j \left(\frac{s_j \ln^2(s_j)}{2} + \beta \varepsilon_j s_j \ln(s_j)\right)$$

The fourth line suggests obviously to identify α and $q$. Notice that the effect of using Tsallis entropy is to make an effective diagonalization, replacing in (12) the sums over the index *i* by a single term. Since these are the sums over the heat bath states, this means that we are replacing the Σ-B interaction with a sort of mean-field approximation. In conclusion, the appearance of non-ortodox entropy appears related just to the fact that we are trying to get correct results using the wrong functional ($F_C$ -Eq. 9-instead of *F*-Eq. 7).

We have shown that it is possible to identify the non-extensivity parameter $q$ with α. But the latter (or, better, 1 - α) is simply a measure of the fluctuations of the heat bath around its most probable state. Therefore $q$ is by no means an intrinsic property of Σ and instead is completely determined by *B*. If no fluctuations are possible, α = 1 and the system Σ is endowed with Shannon entropy. This can happen either because Σ and *B* do not interact at all, so each of them is described within the microcanonical ensemble; or because *B* has a huge number of degrees of freedom (thermodynamic limit) so that its most probable state is overwhemingly probable.

We point out some consequences of this interpretation. Let us consider two systems $\Sigma_1$ and $\Sigma_2$, modelled through non-ordinary statistics with different non-extensivity exponents (i.e. $q_1 \neq q_2$ and both ≠ 1). One question sometimes found in literature is: what is the statistics of the combined system? There does not appear to be a simple way of writing an expression of the kind (1) from two subsystems which, individually, are described by two different non-extensivity exponents. Indeed, there are several papers by Wang and other authors [9] where this issue is debated, without reaching a satisfactory answer. We may mention also the paper [10], where some steps in this directions were made, but without conclusive results (strictly speaking, this paper states that zeroth-law of thermodynamics applies to quasiequilibrium systems with long range interactions. This is not the same as saying that it holds within nonextensive statistical mechanics, although some connections undoubtedly may be seen). We see instead that now, within the present framework, the question itself becomes ill-posed. Infact, either $\Sigma_1$ and $\Sigma_2$ are in contact with the same heat bath, and in this case they are forced to have the same $q$, or they are in contact with two different heat baths $B_1$, $B_2$. But in this latter case the two independent systems to be considered are $(\Sigma_1 + B_1)$, $(\Sigma_2 + B_2)$. These composite systems follow ordinary extensive statistics, hence ordinary addition of independent entropies applies: It is not possible to have two systems interacting with different $q$ indices.

A related issue is: let us suppose to put into thermal contact the two heat baths $B_1$, $B_2$, isolated from the outside. We have therefore the whole system $B = B_1 + B_2$, and $B_1$, $B_2$ could be alternatively regarded as the heat bath or the system studied. Let us suppose that a system $\Sigma$ be endowed with $q_1$ ($q_2$) statistics if it were separately interacting with $B_1$ ($B_2$). Then, there appears an (apparent) contradiction within our scheme: infact, if we chose $B_1$ as heat bath, and $\Sigma = B_2$, then $B_2$ should follow $q_1$-statistics, while, in the opposite case it is $B_1$ to follow $q_2$-statistics. But is this a real contradiction? Not at all. Infact, it is necessary to remind the distinction between thermal bath and system: from an operational point of view, they are distinct by the fact that the former acts on the latter, but no feedback is possible; the state-and the statistics-of B are not appreciably influenced by the presence of $\Sigma$. We must therefore ask how $B_1$, $B_2$ relate to each other. If, qualitatively speaking, one of them-say, $B_1$-is of macroscopical size with respect to the other, we are still in the situation of one heat bath ($B_1$) and a system ($B_2$). Previous equations are not symmetrical with respect to the interchange bath-system: no conclusion about the statistics of $B_1$ can be drawn; indeed, its statistics *is given as known*. Therefore, we simply expect the dynamics of $B_2$ to be distorted to the extent that now it follows $q_1$-statistics. Let us suppose, instead, that $B_1$, $B_2$ are of comparable magnitude and able to appreciably affect each the dynamics of the other. In this case there is no reason to pretend that the unperturbed statistics $q_1$, $q_2$ still hold. Instead, the whole system will follow an "intermediate" statistics. We can still use the heat bath-system picture, but now each of the two systems will be endowed with the same statistics. The intermediate statistics will not likely even be Tsallis-like. This result is important since it is exactly what has been found in the investigation [11], prior to this study, which is therefore an independent confirmation. Also, a different proof of the impossibility of merging two Tsallis distributions into one, which does not make use of entropy, is provided in Appendix.

A comment is in order about the earlier statement according to which finite heat baths would be suitable generators of power-law statistics. Rigorously, this cannot be true: if B is finite, $\Sigma$ and (B + $\Sigma$) of course also are, thus are endowed with a total finite energy. Hence, arbitrarily high velocities are not allowed and the velocity PDF must have a finite support (We refer here, for simplicity, to energy and velocity PDFs. Of course, the same can be said for any other quantity). Indeed, the results shown in this paper rely upon a truncation to first order of a Taylor expansion in powers of (1 - q) of the correct functional. The true statistics arises from the full expression, hence power-law PDFs necessarily are only approximations of the true PDFs. However, differences appear when sampling extreme tails of the PDF, which experimentally is hard to perform. Furthermore, the quality of the approximation is good as long as (1 - q) is a small parameter. This has a number of consequences: 1) first of all, only for finite heat baths characterized by large numbers of degrees of freedom (i.e., q ~ 1) we expect Tsallis-like statistics to yield a good description of reality and, in all cases-again-it must break down for very high velocity. 2) Our formalism can indeed accomodate several nonstandard statistics, all of them parameterizable by the single parameter q (and, of course, it could even be generalized to multiparameterized statistics). However, since q ~ 1 must hold, all the statistics must be very similar between them since they must all approach the common Gaussian limit (q = 1).

Finally, some further comments are in order:

I) in this work we have focussed on Tsallis' NESM since, as we have already told, it is the most famous one. Of course, the above reasoning apply equally well to any non-extensive statistics (see the above paragraph: all of them count among Luzzi et al's Unconventional Statistical Mechanics).

II) Recently, it has been understood that Tsallis statistics is not relevant to equilbria of simple systems but to quasi-equilibria, or non-equilibrium stationary states, of complex systems (see, e.g., [12]). We want to point out that this does not make any difference for the present work. Indeed, regardless of the fact that we are speaking about equilibria or just quasi-equilibria, in the standard formalism of NESM, the generalized entropy is the fundamental quantity from which everything is derived.

**Appendix**

A well known problem with Tsallis' NESM concerns the zeroth law of thermodynamics in presence of two or more systems characterized by different entropic $q$ indices: a precise definition of a common temperature for such systems is still missing. Intensive investigations have been carried on in the past in particular by Wang *et al* to address this issue [9], but only with limited success: indeed, papers [9] are somewhat successful in deriving the zeroth law, but only at the expense of further *ad-hoc* postulates.

In this appendix we provide a fairly simple argument suggesting that approaches *á la* Wang do appear unfruitful: a zeroth law of thermodynamics, apparently, can only be written down for systems with the same index $q$.

As a first step, let us remind the standard form for the probability density function (PDF) in a system ruled by Tsallis statistics, with the "energy" $E$ as independent variable:

$$p(E) = \frac{1}{Z} \frac{1}{(1+\beta(q-1)E)^{\frac{1}{q-1}}} \quad \text{(A1)}$$

We identify three regimes, depending upon the value of $E$: I) first, the "low-energy" region, $\beta(q-1)E \leq 1$. This region is scarcely interesting from our point of view: in this range, regardless of its precise dependence on $E$, $p$ can be expanded into a Taylor series truncated to first order, hence no real information about the true shape of the PDF can be got. II) Then, there is the usual "intermediate" range, $\beta(q-1)E \geq 1$, where the precise analytical form (exponential, stretched exponential, power-law, etc...) for $p(E)$ is looked for. III) Any physical system is, however, finite, hence there must be an extreme-value (very high *E*) region, where *p*(*E*) falls sharply to zero. We will stay away also from the region affected by these "finite-size" effects, and consider only the "intermediate" region (II).

Under this hypothesis, we can discard the unity term in the denominator of (A1) and write

$$p(E)dE \approx \frac{1}{Z}\frac{1}{(\beta(q-1)E)^{\frac{1}{q-1}}}dE \approx \frac{1}{Z\beta(q-1)^{\frac{1}{q-1}}}\exp\left[-X\left(\frac{1}{q-1}-1\right)\right]dX$$
$$\equiv C_q \exp(-BX)dX \quad .$$

(A2)

where we have set $\beta E = \exp(X), B = 1/(q-1)-1$.

Eq. (A2) admits a straightforward interpretation: at least in a restricted energy region (but it is the most important region for this kind of studies), the probability distribution of a system ruled by Tsallis statistics can be mapped into that of a system ruled by ordinary Maxwell-Gibbs statistics, in thermal equilibrium with its surroundings, with "energy" *X* and "temperature" 1/*B*. Notice, however, that writing *p*(*E*) in the form (A2) is not necessary to our purposes, just slightly simplifies the following computations.

Let us now move to the next step, and consider two systems $S_1$, $S_2$, parameterized with the nonextensivity indices $q_1$, $q_2$ respectively. In order to give an intuitive picture of the problem, we may think of each system as a box containing a large number of particles, respectively $N_1$ and $N_2$. Each particle is given a value: its "energy" *E*. Apart for the individual *E* label, particles are identical between them. The statistical distribution of *E* follows Tsallis statistics, with the corresponding value for *q*, in each box. In other words, the probability of randomly picking up a particle labelled with energy E from box *i* (*i* =1,2) is given by Eq. (A1) with $q = q_1, q_2$.

The attempt of describing the combined system ($S_1 + S_2$) in terms of a single distribution is tantamount to pouring all the particles into a single box. The resulting combined statistics will come out by randomly picking a large number of particles out of this box. Let us ask how much the probability is of extracting from the urn a particle labelled with energy *E*. The answer is, trivially:

$$p(S_1 + S_2, E) = \frac{N_1}{N_1 + N_2} p(S_1, E) + \frac{N_2}{N_1 + N_2} p(S_2, E) \quad . \quad (A3)$$

In this expression, $N_i/(N_1+N_2)$ is the probability of choosing a particle out of those previously in the box $i$, and $p(S_i,E)$ is the probability that this particle had energy $E$. Since we are interested to the high-energy tail of the distribution, we can replace $p(S_i,E)$ by its form (A2), which yields:

$$p(S_1 + S_2, X) = \exp(-B_1 X - \mu_1) + \exp(-B_2 X - \mu_2) \quad (A4)$$

where $\mu_i = -\ln(N_i/(N_1+N_2)) - \ln C_{q_i}$ is a sort of chemical potential.

The l.h.s. of (A4) must be, in its turn, a Tsallis distribution, hence, we get the final result

$$\exp(-B_{1+2} X - \mu_{1+2}) = \exp(-B_1 X - \mu_1) + \exp(-B_2 X - \mu_2) \quad (A5)$$

It is just obvious that Eq. (A5) cannot be fulfilled for arbitrary $X$, given generic $B_i$, $\mu_i$, the exception being, as expected, the equal-$q$-case, for which $B_1 = B_2 = B_{1+2}$, $\mu_1 = \mu_2 = 1/\ln(2) \mu_{1+2}$. This concludes our proof.

Other interesting consequences can be drawn from Eq. (A4). In order to fix ideas, let us set $B_1 < B_2$. As long as it does not hold $\mu_1 \gg \mu_2$, the former exponential on the r.h.s of (A4) will prevail over the latter, hence the asymptotic behaviour of the combined distribution will essentially match a $q_1$-statistics. On the other hand, widely differing values between $\mu_1$ and $\mu_2$ can locally override this trend and give birth to an extremely complicated behaviour. Since different $\mu$'s can be made roughly correspond to hugely different numbers of particles (although there is also a $q$-dependent contribution from the partition function, $C_q$), this statement is equivalent to saying anything but that the overall statistics of the system is driven by the larger-in-size of the two subsystems.

Some final comments are: the paper [13] has recently appeared, in which the possibility of defining an unique effective temperature for non-equilibrium critical systems is criticized. These systems are exactly between those though amenable to study using nonextensive statistics. It is also remarkable that the impossibility of defining a common temperature, in that work, is directly related to the existence of fluctuations away from the Gaussian limit.

Secondly, within Beck's superstatistics interpretation of NESM [14], previous results do appear fairly reasonable: since, there, the entropic index of a system is a measure of the fluctuations of some generalized temperature, it does seems difficult to relate within one and the same formulation fluctuations from two completely uncorrelated systems.